**Phonemic evidence reveals interwoven evolution of Chinese dialects**

Meng-Han Zhang[1], Wu-Yun Pan[2], Shi Yan[1]*, Li Jin[3, 4]*

[1]Ministry of Education Key Laboratory of Contemporary Anthropology and Collaborative Innovation Center for Genetics and Development, Human Phenome Institute, School of Life Sciences, Fudan University, Shanghai, 200438, China

[2]Department of Chinese Language and Literature, Fudan University, Shanghai, 200433 China

[3]State Key Laboratory of Genetic Engineering, and Collaborative Innovation Center for Genetics and Development, School of Life Sciences, Fudan University, Shanghai, 200438, China

[4]Chinese Academy of Sciences Key Laboratory of Computational Biology, CAS-MPG Partner Institute for Computational Biology, SIBS, CAS, Shanghai, 200031, China

*Correspondence and requests for materials should be addressed to Shi Yan (yanshi@fudan.edu.cn) and Li Jin (lijin@fudan.edu.cn).

**Abstract**

**Han Chinese experienced substantial population migrations and admixture in history, yet little is known about the evolutionary process of Chinese dialects. Here, we used phylogenetic approaches and admixture inference to explicitly decompose the underlying structure of the diversity of Chinese dialects, based on the total phoneme inventories of 140 dialect samples from seven traditional dialect groups: Mandarin, Wu, Xiang, Gan, Hakka, Min and Yue. We found a north-south gradient of phonemic differences in Chinese dialects induced from historical population migrations. We also quantified extensive horizontal language transfers among these dialects, corresponding to the complicated socio-genetic history in China. We finally identified that the middle latitude dialects of Xiang, Gan and Hakka were formed by admixture with other four dialects. Accordingly, the middle-latitude areas in China were a linguistic melting pot of northern and southern Han populations. Our study provides a detailed phylogenetic and historical context against family-tree model in China.**

**Main Text**

**Introduction**

As a major member of the Sino-Tibetan family, Chinese is one of the most widely spoken languages in the world with 1.3 billion speakers[1]. The earliest concept of Chinese dialects can be traced back to ~2500 years before present in the Zhou Dynasty[2]. Among the 7 modern dialect groups, Mandarin dialects, accounting for 70% of all Chinese-speaking populations, are mainly located in Northern China, while the Wu, Xiang, Gan, Hakka, Yue, and Min dialects are distributed in Southern China. This traditional dialect classification is based on the sound change from Middle Chinese (Chinese spoken at ~ 601 AD), and exhibits extensive differences in different geographic regions. Notably, these dialect groups are barely intelligible among each other, with a diversity comparable to the entire Romanic or Germanic languages. The Chinese writing system does not use letter-based phonograms like English, but instead uses character-based ideograms. In contrast to the sound systems of Chinese dialects, the writing systems of Chinese dialects are rather homogeneous due to the strong influence of a standard written language system. Specifically, when new concepts and words are introduced from other dialects, local populations tend to adopt the same writing forms of the new words, but speak them in their local pronunciation. Hence, the sound systems of Chinese dialects show higher diversity than the writing system, much like the case of Arabic language. In other words, while the lexical systems of Chinese dialects were frequently replaced by the standard language, the phonemic systems retained both local pronunciations and substantial historical signals.

As the fundamental unit of sound systems, phonemes have been utilized to examine language origins and changes[3-5]. Some researchers suggest that in contrast to lexical systems, phoneme inventories may be more conservative and allow for greater insights into the evolution of languages[6]. In this study, by comparing differences in phoneme inventories, we analyzed fine-scale structures among 7 Chinese dialect groups with 20 dialect samples for each dialect group (see Fig. 1a). All of the phoneme inventories we used were compiled by the co-authors of this study and other Chinese historical linguists. To our knowledge, this study presents the first effort by conducting admixture inference to decompose the underlying structure of the diversity of Chinese dialects using phoneme inventories.

**Results**

**Global principal component analysis of phonemic variation in China**

To investigate the relationship among Chinese dialects, we applied Principal Component Analysis (PCA)[7] on presence or absence of phonemes in compiled data from the 140 Chinese dialects. The plot of the first two principal components (PCs) in Fig.1b shows three apices: Mandarin in the North, Min and Yue in the southernmost regions of China, and Wu in the eastern coastal regions.

As a dialect group located in the center of China, Xiang did not become a clear independent cluster due to high variances of the first two PCs, suggesting significant internal phonemic diversity. The Gan dialect samples were primarily between two clusters of Mandarin and Min / Yue in the PC plot. The majority of the Hakka dialect samples we gathered clustered with the Min and Yue dialects, while the rest of the samples were clustered close to Mandarin. Based on Procrustes analysis[8], we identified a significant spatial concordance (Procrustes $t_0 = 0.7137$, $p$-value $< 10^{-5}$) between the first two PCs and geographic locations for the 140 Chinese dialects.

In addition, we found a positive correlation between pairwise phonemic distances of dialect samples and corresponding geographic distances (Mantel $r = 0.2382$, $p$-value $= 1.2988 \times 10^{-7}$). Such a significant correlation revealed that geographically adjacent dialect samples had similar phonemic systems, while distant dialects showed significant differences in their pronunciations. Moreover, major differences among the phonemic inventories of these dialect samples was correlated with the geographic differences in latitudes (PC1 vs Latitude: Spearman's $r = 0.7294$, $p$-value $= 1.6364 \times 10^{-24}$; PC1 vs Longitude: Spearman's $r = 0.1126$, $p$-value $= 0.1853$). Statistical regression at the level of dialect groups further verified a significant association between median PC1 values and latitudes of central locations for each dialect group (Latitude: $R^2 = 0.683$, $p$-value $= 0.022$; Longitude: $R^2 = 0.021$, $p$-value $= 0.757$), as shown in Fig. 1c. These findings revealed a north-south gradient of phonemic differences in Chinese dialects, in agreement with the observations in population genetic studies[9,10] and demographic accounts[2,11,12].

**Neighbour-Net and Delta scores for Chinese dialects**

Neighbour-Net with abundant parallelograms inside illustrated complicated relationship among Chinese dialects (Fig. 1d). The clustering results in such networks were consistent with the PC plot. We used delta scores to measure the complexity of Neighbour-Net. Due to such horizontal influences, the evolution of Chinese dialects does not conform to the family-tree theory[13], but instead conforms to the wave theory[14]. The family-tree theory concerns language evolution induced by social splitting and language divergence, whereas the wave theory emphasizes the importance of horizontal influences in the process of language contact.

In terms of dialect groups, we hypothesized that compared to the average delta score[15] of the given network, the lower delta scores of one dialect group with much samples indicated less horizontal influence. Fig. 2 shows boxplots for delta scores grouped into the seven Chinese dialect groups ordered by north-south geographic distribution. The different dialect groups exhibited rising and falling tendencies, with the highest delta score being in the Gan dialect. The KW test underlined the significant changing tendency of delta scores in Chinese dialects (KW statistic = 62.94, *p*-value < $10^{-4}$). The results of the KW test suggested that the different Chinese dialects experienced different degree of horizontal influence concerning their phonological systems. In addition, we performed Student's *t*-test to compare the distribution of delta scores for each Chinese dialect group with the overall delta score of the Neighbor-Net (0.3473) (Table 1). The statistical results showed that the average delta scores of the Gan and Hakka dialects were significantly greater than the overall score of the network structure. This result strongly indicated that both the Gan and Hakka dialects experienced more horizontal influence than the other dialects.

Moreover, Gan dialect with the highest average delta score could divide the changing tendency into two parts (Fig. 2). The left portion showed Xiang had a slightly higher delta score than Mandarin and Wu, but the difference was not significant (KW statistic = 1.853, *p*-value = 0.3959). A sharp increase between Xiang and Gan indicated that a strong degree of horizontal influence occurred in the Gan dialect. Based on the demographic history of China, this degree of horizontal influence could be considered as the result of the strong influence of long-distance population migration and cultural diffusion from the dialectal areas of Mandarin, Wu and Xiang to the speaking area of Gan dialect. In the other part, there was a significant decrease in the delta scores of Hakka, Min and Yue (KW statistic = 23.28, *p*-value < 0.0001). This decrease indicated that the effects of horizontal influence decreased between the Gan speaking area and areas father to the south.

**The fine-scale structure of Chinese dialects**

To delineate the admixed structure of the dialects, we performed admixture inference approaches. These analytical approaches have been successfully used to estimate mixing proportions of population admixture in anthropology and population genetics[16,17]. To further decompose the underlying structure of the diversity of Chinese dialects, we applied STRUCTURE program[18] to the complete phoneme data (see Fig. 3 and Supplementary Table S1). At K=2, we observed a north-south division for Chinese dialects, similar to the findings based on lexicons and grammar[11,19]. We determined that Gan and Hakka were two mixed languages that experienced strong horizontal influence from the northern (Mandarin, Wu and Xiang) and southern (Min and Yue) dialects. Gan had a closer link to the northern dialects (north: 55.7%, south: 44.3%), whereas Hakka was closer to the southern dialects (north: 24.4%, south: 75.6%). At optimal K=3, we identified the Wu dialect as an independent component corresponding to central dialect. At K=4, we observed a new component introduced primarily in the Gan and Hakka dialects. Based on historical linguists' findings, we inferred that the component was most likely an archaic phonological one of Old Southern Chinese[11] or the Tai and Miao-Yao substratum[20].

Moreover, Xiang had obvious admixture signals from the Mandarin and Wu dialects; however, the causes that produced the admixture signals were different. The Mandarin influence produced phoneme transmissions in the phonological system of Xiang[11], while shared phoneme retention (such as voiced plosives and affricatives) resulted in the observed common component in Wu and Xiang. These voiced consonants were inherited from Middle Chinese in 300–1100 AD, although they have been dropped in most modern Chinese dialects[21]. In addition, 3-population test[22] ($f_3$) produced admixture signals in the Gan, Hakka and Xiang dialect groups, since each of them had at least one negative value of the $f_3$ statistic when given two arbitrary dialect groups as sources of a potential mixed language.

To further determine explicit sources for the three mixed languages, we adopted an alternative admixture inference algorithm implemented in *MixMapper*[23]. We reconstructed phylogenetic relationships among the Chinese dialects using two-way mixing models, in which we assumed that each admixed language had two potential sources. As shown in Fig. 4, the Xiang dialect group was primarily influenced by Mandarin and Yue. Meanwhile, we determined that the Gan dialect was a mixed dialect of Mandarin and Yue. We interpreted the multiple contributing sources of Hakka as Yue (46.80%) and Wu (53.20%), or Yue (59.2%) and Mandarin (40.8%) (see Supplementary Table S2). The results were supported by demic evidence on the demographic migrations from Northern China to Southern China which changed the local population structures and even culture[9-11,24].

**Discussion**

In this paper, we introduced population genetic methods to delineate the phoneme inventories of Chinese dialects. Based on admixture analyses, we decomposed the fine-scale formations of Chinese dialects. Historically, massive demographic movements among Chinese populations who spoke different dialects always produced substantial language contact. According to historical records, there were three important waves of southward migrations in China. The first migration wave occurred during the Western Jin Dynasty (265–316 AD); the second wave occurred during the Tang Dynasty (618–907 AD); and the third wave took place during the Southern Song Dynasty (1127–1279 AD)[24]. The northern immigrants caused significant and continuous changes in the genetic makeup of populations in southern China[9], especially the exceptionally large migrations that took place during the Tang Dynasty. Apart from genetics, southward migrations and dialectal dispersal resulted in substantial language contact between northern and southern dialects, including contact between northern dialects and ethnic minority languages spoken in southern China[20]. This degree of language contact was crucial to dialectal formation and change in China[11].

Different levels of language contact result in varying degrees of horizontal language influence[6]. Frequent or deep language contact often systematically changes intrinsic properties of languages (e.g. phoneme inventory), and even produces language admixture. In this study, we considered language admixture as a case of deep horizontal influence. Therefore, we considered the interwoven evolutionary scenario of Chinese dialects as being dominated by the complicated socio-genetic situation in China such as the multiple waves of centrifugal populations and cultural dispersal[2,19]. Notably, the geographic locations of Xiang, Gan and Hakka are a cluster surrounded by other Chinese dialects in south central China. Therefore, these regions could be inferred as a linguistic melting pot of northern and southern Han populations, consistent with other demographic evidence[11,25].

In addition, different aspects of language systems, such as phonemic systems and lexical systems, experience their own evolutionary processes shaped by linguistic[26], social[27] and ecological factors[28]. Linguists have found out that a single phoneme can be transmitted vertically and horizontally among different languages[26,29,30], and can change over time within a language following the rules of natural sound change[26,30]. A holistic phoneme inventory, however, evolves slowly under mutual phonological constraints among distinct phonemes at the systematic level[5,6]. We summarized that the evolution of phoneme inventory (e.g. phonemic transmission) should be a conservative process that relies on three prerequisite conditions.

The first prerequisite is time duration. It is easy for individual to borrow a word or learn a sentence structure from other language more or less immediately. However, the systematic changes of languages require longer time duration because they embody morphological changes (e.g. affix loss)[31], phonological shifts (e.g. sound chain shifts)[32], or syntactical optimization (e.g. dependency length minimization)[33]. In addition, such linguistic changes also necessitate sufficient time duration to reach an agreement in a finite speaker community.

The second prerequisite is population size. Languages with large populations of speakers are typically considered to be more prestigious language than the various other languages spoken in the geographic vicinity. Hence, systematic changes in language are embodied in minority languages as they attempt to learn a prestigious language. Eventually, minority language speakers either achieve a successful language shift or become bilinguals. However, colonization is a special case because colonists usually have a small population size. Due to the prestige of colonizers' culture and language, colonial people can easily transfer to bilingual status and achieve language shift at lexical level, and even at the phonological level. In a word, the basic facts of language acquisition observed at the population level must have the significance in the evolution of phonological systems.

The third prerequisite condition is the degree of language or cultural contact. A large degree of such contact induced by substantial demographic activities can gradually result in linguistic and cultural convergence. On the one hand, nearby mass population introgression can lead to convergent evolution of geographically close languages[30,34-37]. On the other hand, continual long-term population migration can reduce the differences in linguistic features between two unrelated languages. Regardless of convergent evolution or the reduction of differences between languages, such cases only reflect such changes at the population level, not the individual level.

Based on the above three prerequisite conditions, if a series of phoneme transmissions between two languages are detected, there is reason to believe that these two languages have experienced diachronically long-term, deep contact with each other. Such deep contact also provides an opportunity to develop bilingualisms[38]. Thus, the structure of phonemic systems preserves several conservative phonetic features, which enable us to reconstruct ancient language relationships and track the historical trajectory of language evolution[39].

However, under very strong cultural pressures does the structure of a phoneme inventory tend to become destabilized, such as gaining or losing phonemic contrasts[6]. In addition, phoneme inventories can be affected by substantial population processes[4,26] because drastic changes in population structure can produce strong cultural pressures and rapid language changes. Accordingly, China, with its long and complicated history, is a satisfactory case for the study of demographic migration and cultural diffusion.

While determining complex language history still remains a considerable challenge, our work provides an encouraging alternative perspective for the study of language contact and admixture at both method and data levels. In addition, our work shows a major attempt to fulfill an attractive piece of the jigsaw puzzle about language evolution in the history of Chinese culture.

**Materials and Methods:**

**Phoneme dataset of the Chinese Dialects**

We used the phoneme dataset of 140 Chinese dialects compiled by co-author, Wu-Yun Pan. The dataset is a subset of East-Asian language database copyrighted by Fudan University (URL: http://ccdc.fudan.edu.cn/bases/index.jsp). The dataset contained all consonant and vowel systems of the 140 Chinese dialect samples. These samples can be linguistically classified into seven major Chinese dialects, namely Mandarin, Wu, Xiang, Gan, Hakka, Min, and Yue. Each dialect had 20 samples. We collected total 100 distinctive phonemes, including 70 types of consonants and 30 types of vowels in the phoneme inventories of the Chinese dialect samples. We then collated the presence/absence matrix of these phonemes (see Table S4).

To facilitate comparisons between phonemic inventories of these dialects, we should standardize the representations of phonemes throughout our dataset. We made the following processing of the dataset:

(a). as a variation of [n], [ŋ] generally appears in front of mid and high vowels such as [i]. However, in most Chinese dialects, there is a significant perceptual difference between [n] and [ŋ], but sometimes not between [n] and [l] (e.g. in Wǔhàn). In other words, there is phonemic distinction between [n] and [ŋ]. Therefore, we remained these two different phonemic symbols in our data.

(b). we standardized the zero consonant (or zero initial) in front of [u] as [w]. Similarly, the zero consonant in front of [i] was standardized as [j].

(c). in some dialects (e.g. Tǎiníng), [h] and [x] are phonemic contrast due to sources of archaic Chinese Phonology. However, there is no phonemic contrast between [h] and [x] in some Chinese dialects. In other words, [h] and [x] are allophones with each other. Therefore, we standardized these two phonemes into [h]. In the same way, we used [h] to represent [ɣ].

(d). the vowels we collected in our study consisted of monophthongs and the primary vowels extracting from diphthongs, and syllable structures of Consonant-Vowel (CV) and Consonant-Vowel-Consonant (CVC).

(e). we standardized [a] and [ɑ] into [a] when there was no phonemic contrast between these two vowels; if there was only [ɑ] but not [a] in the phonemic inventory of one dialect, we still standardized [ɑ] as [a].

**Principal Component Analysis and Procrustes Analysis**

For phonemic data of the 140 Chinese dialect samples, we performed Principal Component Analysis[40] (PCA) on the binary matrix of the phonemes, along with Procrustes analysis[41] of the phonemic PCs versus the geographic coordinates of the dialect samples analyzed. Procrustes analysis[41-43] is a multivariate analysis approach to identify relationships between two or more statistical maps such as genetic-geographic[44] or population-genetic-geographic maps[45]. The rationale of Procrustes analysis is to find an optimal transformation for two or more maps that maximizes the measure of the similarity of the transformed maps, and to then score the similarity between the two optimally transformed maps. A permutation test can then measure the significance that a randomly chosen permutation of the points in any one map produces a greater similarity score than that observed for the actual points in the other map[46].

Following Wang et al.[8], we compared two-dimensional PCA maps on the basis of the presence or absence phonemes in the 140 Chinese dialect samples to a geographic map of corresponding samples. We calculated a similarity score based on the statistic $t_0 = \sqrt{(1-D)}$, where D is the minimized sum of the squared distances after Procrustes analysis. We then calculated the empirical *p*-value for $t_0$ values over $10^5$ permutations of the geographic locations. We implemented all computational procedures of PCA, Procrustes analysis and permutation test in Matlab® R2015b (MathWorks, Inc.).

**Correlation between geographic and phonemic distance matrices**

We calculated the pairwise great-circle distances (Orthodromic distances) of 140 dialect samples for the metric of the geographic distance. We then transferred the great-circle distance (*d*) into a logarithm following the formula $log_{10}(d)$. Following the same distance measurement in Creanza *et al.*, we calculated the pairwise phonemic distance matrix based on *Hamming* distance method[47]. We used Mantel test[48,49] to calculate the Spearman's correlation coefficient between the two distance matrices. We performed Mantel test using the Matlab script programed by Enrico Glerean (URL: http://becs.aalto.fi/~eglerean/permutations.html). We set the number of permutations in Mantel test at 10,000.

**Neighbor-Net and delta score**

Some scholars have pointed out that the phylogenetic tree model cannot accurately depict linguistic and cultural evolution because of language contact. As an alternative, some scholars advocate for the network model[50]. The primary advantage of the network model is that it can detect horizontal transmissions in language contact. The Neighbor-Net algorithm is widely used to visualize conflicting signals against tree-like models[51,52], when attempting to establish phylogenetic networks. In this study, we here established a Neighbor-Net using the *Hamming* distance method[47] based on total 100 phonemes from the 140 Chinese dialect samples.

To measure the degrees of horizontal transmission, we calculated the individual delta score for each dialect sample, and an average score for the total 140 samples in the given Neighbor-Net. Delta score was first proposed by Holland *et al.*[15], and has been proven as an valuable parameter for measuring the complexity of reticulations in a phylogenetic structure[53-55]. The value of delta score ranging between 0 and 1 indicates that the relationships between taxa include both tree-like and net-like components[53]. Delta score equals zero when the phylogeny fits a tree-like structure, and one when phylogeny fits a network-like structure. Delta score has been proven useful and effective in other studies of linguistic and cultural evolution to quantify reticulation in data. Furthermore, we statistically compared the distributions of delta scores across different Chinese dialects using Kruskal-Wallis test[56] (hereafter shortened as KW test) in Matlab 2015b. We performed the entire procedure of graphic visualization and calculation in SplitsTree4 (http://www.splitstree.org/) using default settings.

**Structure analysis**

We used the STRUCTURE program[57-59] to analyze the potential mixing components of each Chinese dialect sample under the influence of horizontal transmissions. We conceptually regarded the presence or absence of total 100 phonemes as haploid alleles in phonemes to be correlated with each other. We set the running iteration of the STRUCTURE program to 20,000 iterations, including 10,000 burn-ins. We also ran the STRUCTURE program from K=2 to K=8, with 10 repeats for each K. We then used CLUMPP program[60] to align the repetitions for each K obtained from STRUCTURE. The plot shown in Fig.2 represents the optimal alignments among the 10,000 input orders obtained from a large K-greedy algorithm by using CLUMPP.

To obtain the optimal cluster number that best fit our data, we re-processed the STRUCTURE results using STRUCTURE HARVESTER[61] program (online URL: http://taylor0.biology.ucla.edu/structureHarvester/#) to obtain the optimal number of clusters. The STRUCTURE HARVESTER program provides a fast way to assess and visualize likelihood values across multiple values of K. The process of finding the optimal cluster is based on the Evanno method[59], which performs an *ad hoc* statistic Delta K based on the rate of change in the log probability of data between successive K values. The optimal number K has the largest Delta K.

**Explicit admixture inference**

To further investigate the explicit components of horizontal transmission in language admixture, we applied an alternative inference approach implemented in *MixMapper* v2.0 software[23]. *MixMapper* is a tool for building phylogenetic models of population relationships that incorporate the possibility of admixture[17]. This software can also be thought of as a generalization of the *qpgraph* and *Treemix* programs[62]. The procedures of using *MixMapper* can be divided into two major phases. The first phase is to establish a scaffold Neighbor-Joining tree[63] of un-admixed populations from the $f_2$ distance matrix. Whether the population is admixed or not is determined by 3-population test ($f_3$ test)[22,62], in which the $f_3$ statistic values of un-admixed populations are positive while of strongly admixed populations are negative[64]. In the second phase, *MixMapper* tries to expand the scaffold Neighbor-Joining tree in consideration of fitting an admixed population between pairs of branches on the tree. In such cases, *MixMapper* produces an ensemble of predictions via bootstrap resampling, enabling confidence estimation for inferred results[65]. In this study, we performed 500 bootstraps replicated in *MixMapper* analysis. Noting that, we regarded the seven Chinese dialects as seven language populations with each containing 20 individual samples. Compared to STRUCTURE, the admixture results of *MixMapper* were more explicit at the population level.

**Acknowledgements:** This study is supported by projects at the National Natural Science Foundation of China (31501010, 31401060, and 31521003), the Special Program for Key Basic Research of the Ministry of Science and Technology, China (2015FY111700), the Postdoctoral Science Foundation of China (2015M570316 and 2015T80394), and the Science and Technology Committee of Shanghai Municipality (16JC1400500).

**Author Contributions:** M.Z., W.P., S.Y. and L.J. designed the research structure of this study; M.Z., W.P., and S.Y. performed the research; M.Z., W.P., S.Y., and L.J. analyzed the results; and M.Z., S.Y., and L.J. wrote the paper. The authors declare no conflicts of interest.

**Competing Financial Interests:** The authors declare no competing financial interests.

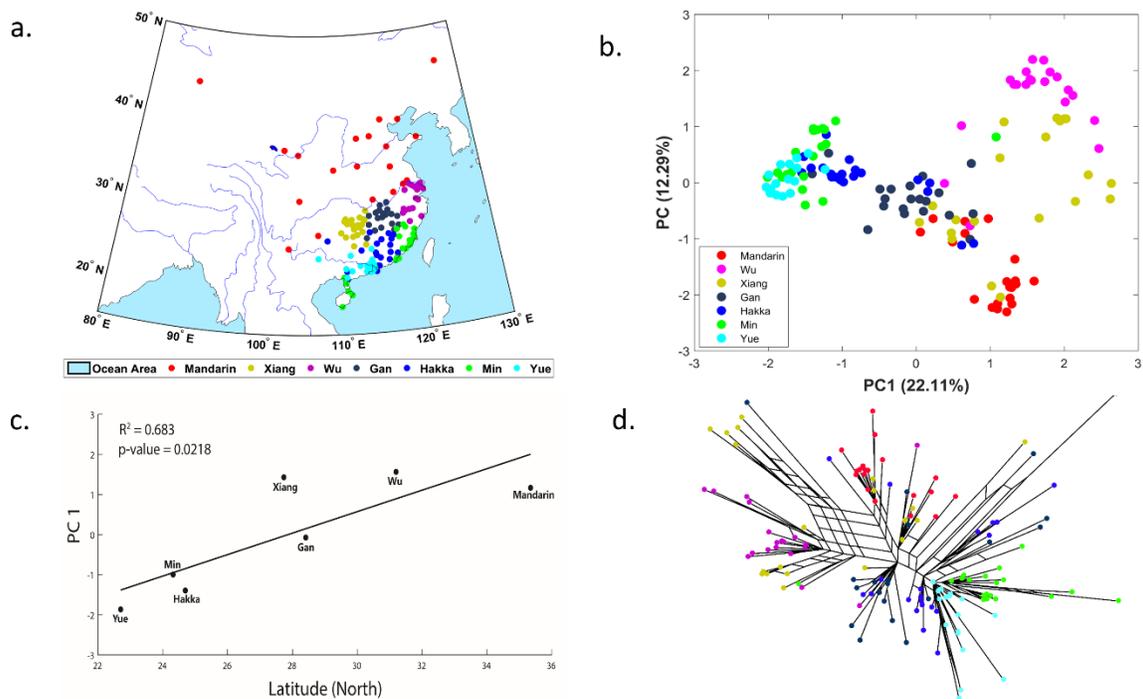

**Fig. 1.** (a) The geographic distribution of the 140 dialect samples grouped in seven Chinese dialect groups (Mandarin, Xiang, Wu, Gan, Hakka, Min and Yue). Different colors represent different dialect groups. (b) The Principal Components Analysis (PCA) of the phonemic systems of the 140 Chinese dialect samples, using the first two principal components (PCs) and their explained percent variances. (c) The correlation analysis of PC1 and the latitude. The *x*-axis represents the geographic centers of each dialect group, indicated by median latitudes. The *y*-axis represents the median PC1 values of each dialect group. The line in the plot shows the regression line ($y = 0.2677x-7.4575$). (d) The Neighbor-Net for the 140 Chinese dialect samples calculated using the Hamming distance matrix. Different colors indicate various dialect groups.

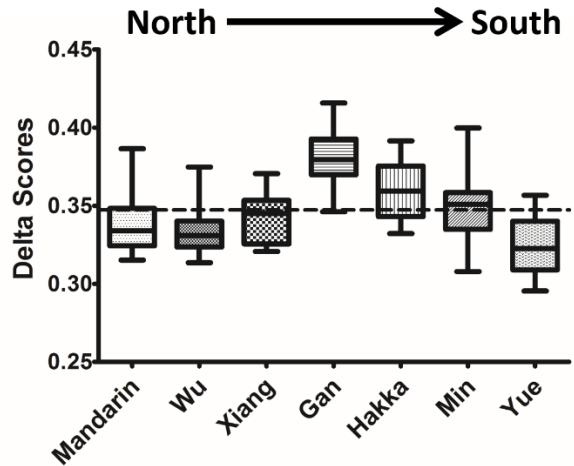

**Fig. 2.** The boxplots with delta scores of seven Chinese dialect groups are displayed. Each box contains 20 dialect samples. The dash line represents the average delta score (0.3473) of all the dialects in Neighbor-Net. The dialect names of axis X are sorted by geographic locations from north to south.

**Table 1.** The statistical summary of delta scores of seven Chinese dialect groups containing 20 samples each. Right-tailed Student's *t*-test is applied to test the difference between each Chinese dialect group and the average delta score of Neighbor-Net (0.3473). Significant *p*-values (*p*<0.05) are shaded in gray.

| Dialect group | Samples | Mean | Std. Deviation | Lower 95% CI | Upper 95% CI | *p*-value |
|---|---|---|---|---|---|---|
| Mandarin | 20 | 0.3398 | 0.0203 | 0.3304 | 0.3493 | 0.9417 |
| Wu | 20 | 0.3353 | 0.0160 | 0.3278 | 0.3428 | 0.8914 |
| Xiang | 20 | 0.3425 | 0.0170 | 0.3345 | 0.3504 | 0.9983 |
| Gan | 20 | 0.3806 | 0.0164 | 0.3729 | 0.3882 | $1.2 \times 10^{-8}$ |
| Hakka | 20 | 0.3594 | 0.0164 | 0.3518 | 0.3671 | 0.0018 |
| Min | 20 | 0.3487 | 0.0213 | 0.3387 | 0.3587 | 0.3834 |
| Yue | 20 | 0.3244 | 0.0194 | 0.3154 | 0.3335 | 1.0000 |

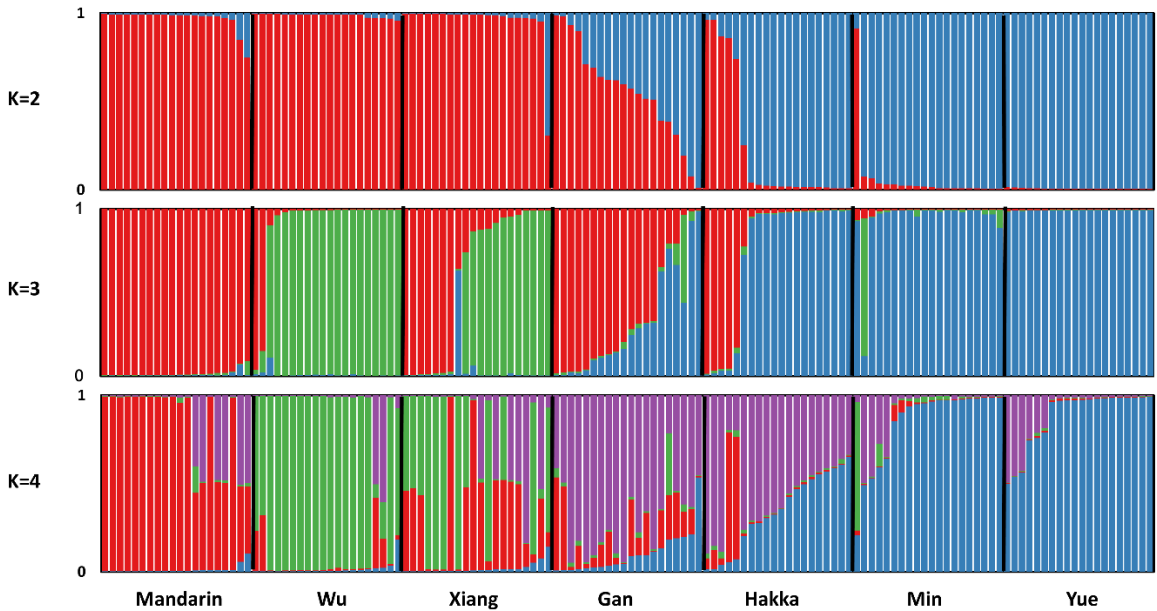

**Fig. 3.** The summary plots of the individual admixture proportions of the 140 dialect samples grouped into seven Chinese dialect groups from K=2 to K=4. The optimal cluster is K=2. Each individual dialect sample is represented by a single vertical bar broken into coloured segments, with the length of the segment proportional to each inferred clusters. The dialects samples were sorted in each dialect group using admixture proportions.

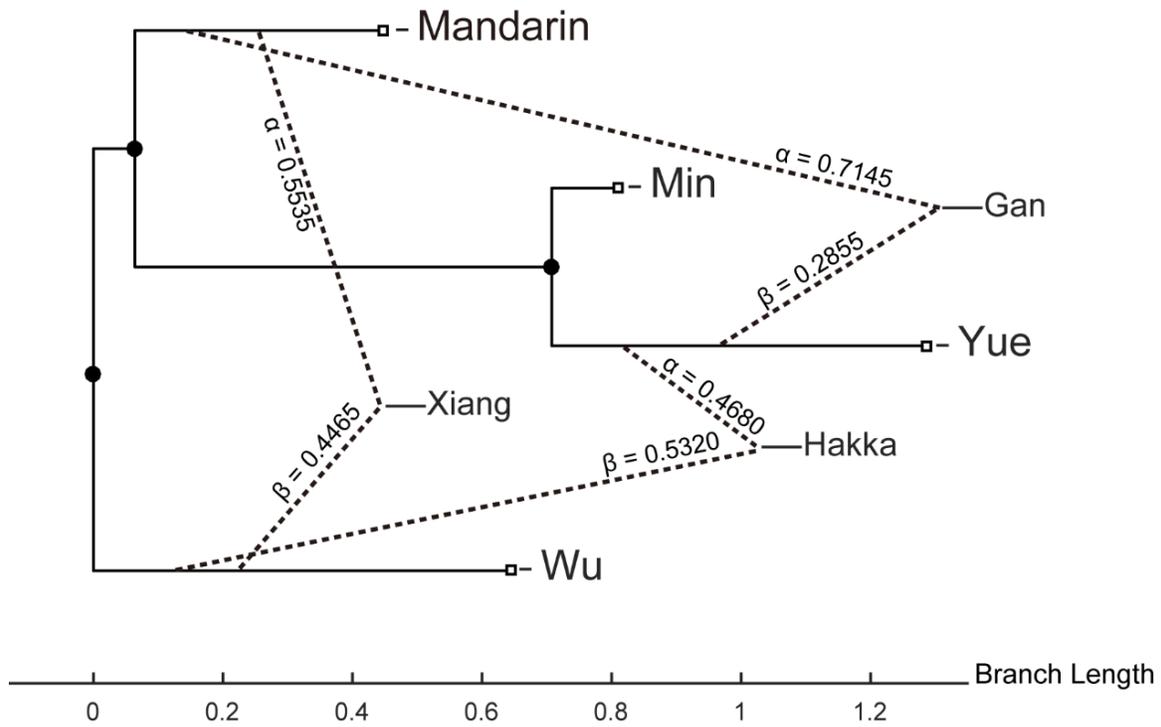

**Fig. 4.** The admixture graphs fitted by first optimizing a scaffold tree consisting of Mandarin, Wu, Min, and Yue. The dashed lines represent the contributors of admixed dialects based on a two-way mixing model. The average admixture proportions are shown on the dashed lines, and 95%CIs of each dashed branch are shown in Table S2. For the Hakka dialect, we only show the contributions of Wu and Yue dialects because they had the greatest number of bootstrap replicates. $\alpha$ and $\beta$ are the proportions of ancestry from Branch 1 and Branch 2, respectively, where $\alpha + \beta = 1$.

**Supplementary information:**

Table S1-S2

Extended Data Table S3

# Supplementary information

**Phonemic evidence reveals interwoven evolution of Chinese dialects**

Meng-Han Zhang, Wu-Yun Pan, Shi Yan, Li Jin

**Sections**

**Table S1-S2**

**Extended Data Table S3**

**Table S1.** The summary of Structure Harvester based on Evanno *et al.*'s method[59]. The optimal cluster (K=3) had a largest value of Delta K.

| K | Reps | Mean LnP(K) | Stdev LnP(K) | Ln'(K) | |Ln"(K)| | Delta K |
|---|---|---|---|---|---|---|
| 2 | 10 | -3285.9200 | 1.5281 | — | — | — |
| 3 | 10 | -2941.3700 | 5.2417 | 344.5500 | 168.9100 | 32.2241 |
| 4 | 10 | -2765.7300 | 21.4521 | 175.6400 | 12.7900 | 0.5962 |
| 5 | 10 | -2602.8800 | 16.4235 | 162.8500 | 82.5200 | 5.0245 |
| 6 | 10 | -2522.5500 | 9.4852 | 80.3300 | 15.3800 | 1.6215 |
| 7 | 10 | -2457.6000 | 20.4437 | 64.9500 | 121.5400 | 5.9451 |
| 8 | 10 | -2514.1900 | 271.4606 | -56.5900 | — | — |

**Table S2.** Mixture parameters for three admixed dialects modeled as two-way admixtures implemented in *MixMapper*. Branch choices are shown that typologies occur for at least 100 of 500 bootstrap replicates.

| Admixed Pop | Branch1 + Branch2[a] | Rep[b] | alpha[c] | Branch1 Loc[d] | Branch2 Loc[d] |
|---|---|---|---|---|---|
| Xiang | Mandarin + Wu | 497 | 0.345-0.762 | 0.234-0.549 / 0.549 | 0.246-0.640 / 0.640 |
| Gan | Mandarin + Yue | 361 | 0.576-0.853 | 0.012-0.340 / 0.578 | 0.382-0.603 / 0.603 |
| Hakka | Yue + Wu | 222 | 0.283-0.653 | 0.175-0.542 / 0.542 | 0.006-0.486 / 0.646 |
| | Yue + Mandarin | 209 | 0.350-0.834 | 0.163-0.614 / 0.614 | 0.016-0.562 / 0.562 |

[a] Optimal split points for mixing populations.

[b] Number of bootstrap replicates (out of 500) placing the mixture between Branch1 and Branch2; topologies are shown that that occur for at least 100 of 500 replicates.

[c] Proportion of ancestry from Branch1 (95% bootstrap confidence interval).

[d] Points at which mixing populations split from their branches (expressed as confidence interval for split point/branch total).

**Extended Data Table S3**

Description: The dataset of 140 Chinese dialect samples including corresponding locations and geographic coordinates (Latitude and Longitude), ISO639-3 (ISO639-6 for Min dialect) and binary coded phoneme types.